\documentclass[a4paper]{jpconf}
\usepackage{graphicx}
\begin{document}
\title{Results on angular correlations with ALICE}

\author{Panos Christakoglou (for the ALICE Collaboration)}

\address{Nikhef, Science Park 105, 1098 XG Amsterdam, The Netherlands}

\ead{Panos.Christakoglou@nikhef.nl}

\vspace{-0.4 cm}
\begin{abstract}
Angular correlations are a sensitive probe of the transport properties of the system produced in nucleus--nucleus 
collisions. Similar studies performed in p--Pb collisions have recently revealed intriguing features as well. In this 
article, we review the latest results on charged and identified particle correlations obtained with the ALICE 
detector at the LHC in both Pb--Pb and p--Pb events.
\end{abstract}

\vspace{-1.3cm}
\section{Introduction}
Angular correlations have emerged as a powerful tool to study the properties of the strongly interacting medium 
created in high energy nucleus--nucleus collisions \cite{Ref:Heinz}. Measurements of azimuthal correlations established 
the picture of a system that behaves as an almost perfect fluid with low value of dissipation i.e. low value of shear 
viscosity. The flow harmonics spectrum revealed that elliptic flow ($v_2$) is the dominant component, 
reflecting the elliptic shape of the nuclear overlap region and the hydrodynamic response of the medium 
\cite{Ref:Alicev2}. Recent experimental results revealed that initial state fluctuations can generate higher order 
flow components, the study of which can further constrain the values of viscosity \cite{Ref:LhcHigherHarmonics}. 
In parallel, the study of high transverse momentum ($p_{\rm{T}}$) probes has also established the picture of a system that is dense and opaque \cite{Ref:LhcHighPt}. 
In this article, I review the latest experimental results on angular correlations from the analysis of Pb--Pb and 
p--Pb collisions at $\sqrt{s_{NN}} = 2.76$ and $\sqrt{s_{NN}} = 5.02$~TeV, respectively, recorded with the 
ALICE detector at the LHC \cite{Ref:AliceJinst}.

\vspace{-0.4 cm}
\section{Anisotropic flow results in Pb--Pb and p--Pb}
\begin{figure}[h]
\begin{minipage}{15pc}
\includegraphics[width=15pc]{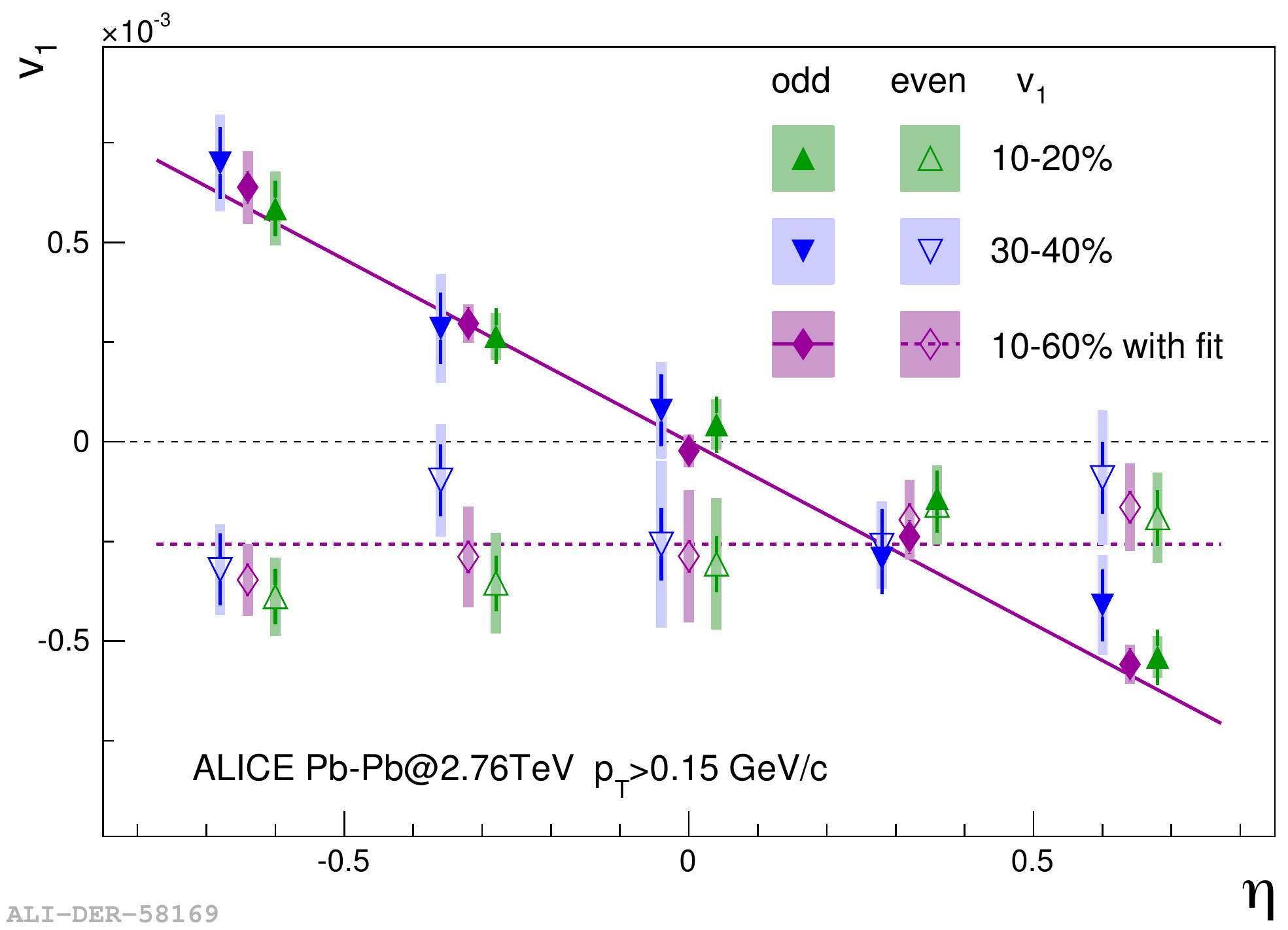}
\caption{\label{fig1}The $\eta$ dependence of $v_1^{odd}$ and $v_1^{even}$ for different centrality class of Pb--Pb collisions~\cite{Ref:Alicev1}.}
\end{minipage}\hspace{2pc}%
\begin{minipage}{17pc}
\includegraphics[width=17pc]{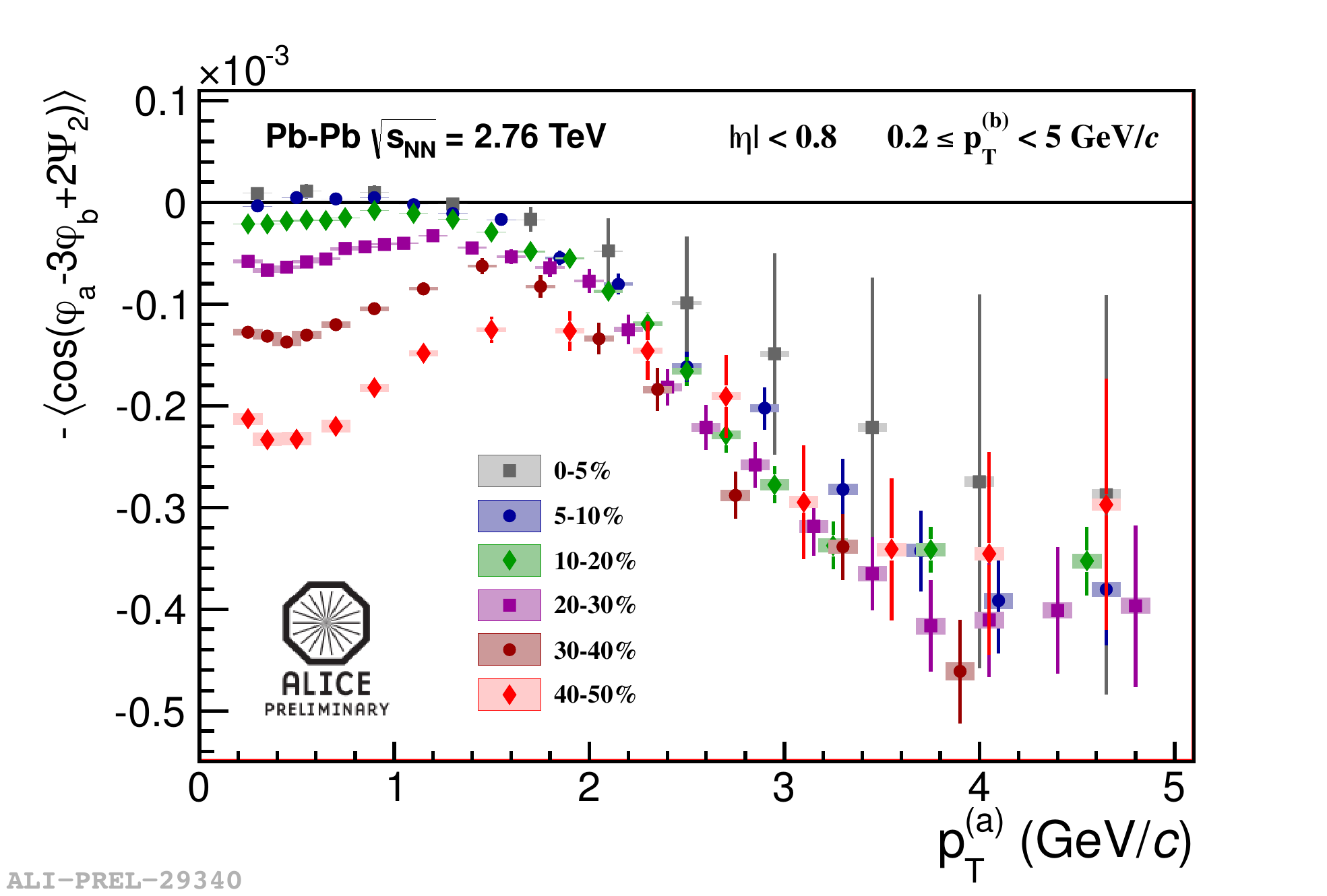}
\caption{\label{fig2}The $p_{\rm{T}}$ dependence of $\langle \cos(\varphi_a - 3\varphi_b + 2\Psi_2) \rangle$, probing the 3--plane correlations, for different centrality class of Pb--Pb collisions.}
\end{minipage} 
\end{figure}

Anisotropic flow studies are based on the description of the azimuthal distribution of particles by a Fourier 
series $dN/d\varphi \approx 1 + \sum_{n=1}^{\infty} 2 v_n \cos[n(\varphi - \Psi_n)]$ \cite{Ref:Voloshin}, where 
$\varphi$ is the azimuthal angle of particles, and $v_n$ are the different flow coefficients developing relative 
to the symmetry planes $\Psi_n$ for each harmonic $n$. The first harmonic $v_1$, the directed flow, affected 
by the amount of nuclear stopping and the initial position of nucleons in the transverse plane, is a sensitive 
probe of the initial conditions of the system. It has been studied in ALICE relative to the deflection of the spectator 
nucleons, reconstructed with the Zero Degree Calorimeters (ZDC) \cite{Ref:AliceJinst} in Pb--Pb collisions at 
$\sqrt{s_{NN}} = 2.76$~TeV. Figure \ref{fig1} presents the pseudo--rapidity dependence of the odd and even 
components of $v_1$ for various centrality classes \cite{Ref:Alicev1}. The former, that originates from flow 
developing relative to the symmetry plane, exhibits a negative slope with little centrality dependence and a 
magnitude that is smaller by a factor of 3 compared to the highest (i.e. $\sqrt{s_{NN}} = 200$~GeV) RHIC energy \cite{Ref:Rhicv1}. On the other 
hand, $v_1^{even}$ has a negative sign, and shows little or no dependence on $\eta$ and on centrality within 
the current uncertainties, consistent with the expectations for a component  originating from fluctuations of the 
initial energy density in the collision.

It is now established, that these initial fluctuations lead to non--zero higher order harmonics that develop relative 
to their own symmetry plane. A step forward in understanding the nature of these fluctuations is to investigate 
whether they are imprinted in the correlation between the different $\Psi_n$. Figure \ref{fig2} presents the $p_{\rm{T}}$ 
differential mixed harmonic correlator $\langle \cos(\varphi_a - 3\varphi_b + 2\Psi_2) \rangle$, where $\varphi_{a,b}$ are the azimuthal angles of particles $a$ and $b$, that probes 
the correlations between the first ($\Psi_1$), second ($\Psi_2$), and third ($\Psi_3$) order symmetry planes. 
The data points for all centralities, represented by different colours and symbols, exhibit non--zero 
correlations, indicating the existence of a 3--plane correlation. It is interesting to note that these results are 
in qualitative agreement with ideal hydrodynamical calculations \cite{Ref:TeanyYan} at low $p_{\rm{T}}$, 
while the same calculations fail to describe the evolution at high $p_{\rm{T}}$.

\begin{figure}[h]
\begin{minipage}{15pc}
\includegraphics[width=15pc]{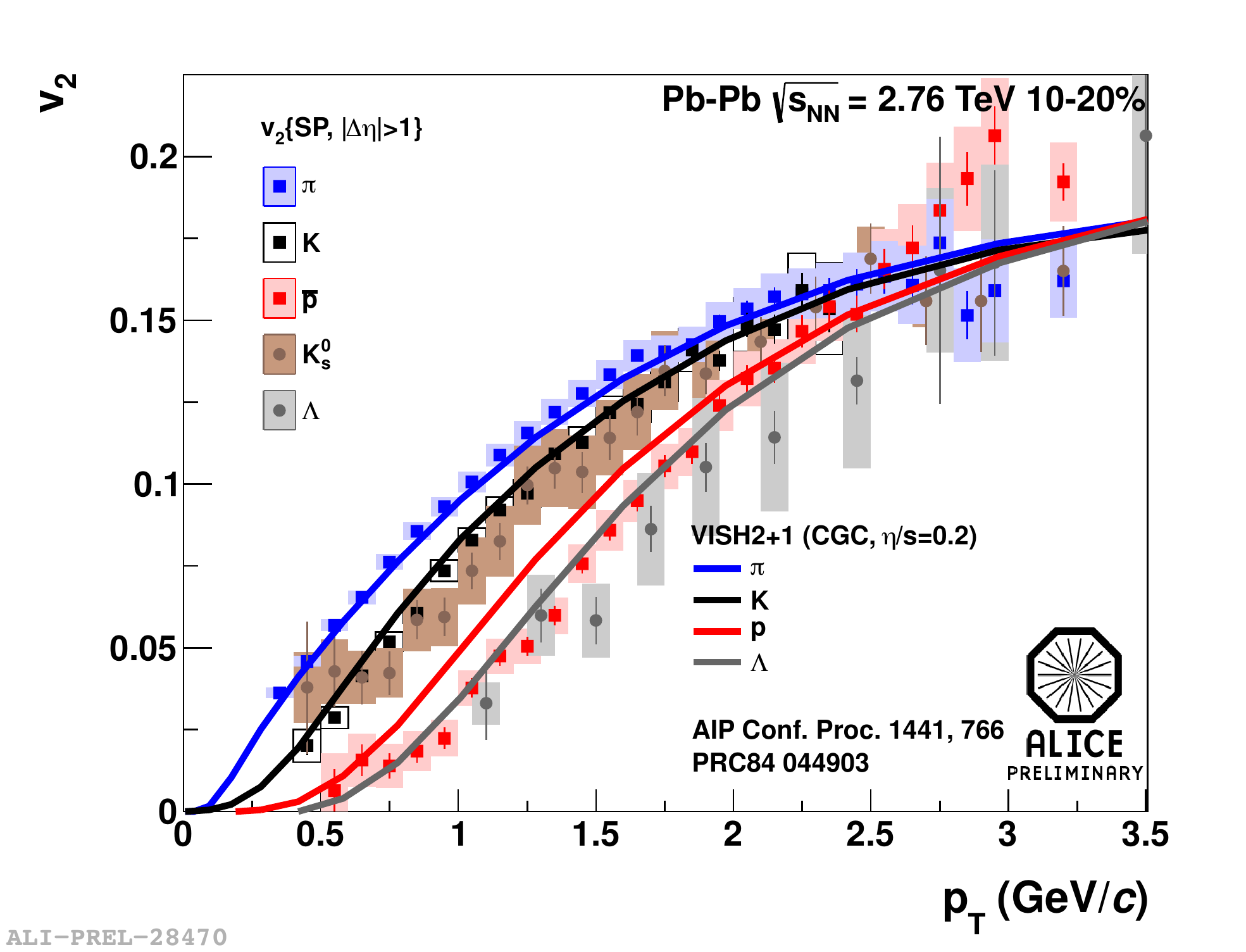}
\caption{\label{fig3}The $p_{\rm{T}}$--differential $v_2$ measurement for central Pb--Pb collisions and for different particle species.}
\end{minipage}\hspace{2pc}%
\begin{minipage}{17pc}
\includegraphics[width=17pc]{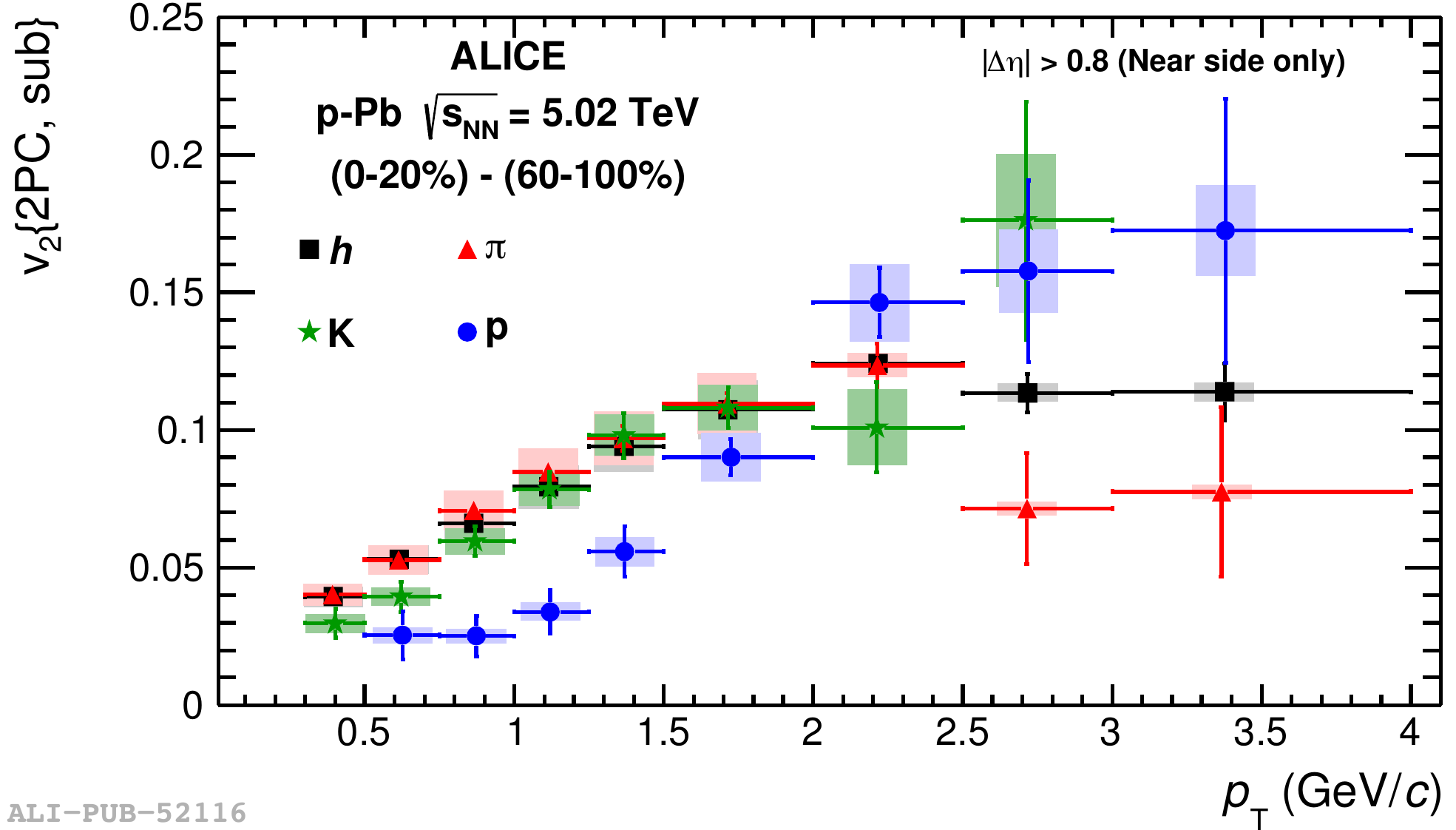}
\caption{\label{fig4}The $p_{\rm{T}}$--differential $v_2$ measurement, after applying the subtraction procedure (see text for details), for p--Pb collisions and for different particle species~\cite{Ref:AlicepPb}.}
\end{minipage} 
\end{figure}

An additional constraint to the value of shear viscosity can be given by studying the second Fourier coefficient, 
the elliptic flow $v_2$, as a function of centrality and $p_{\rm{T}}$ for different particle species. Hydrodynamic 
model calculations predict a characteristic dependence of $v_2$ on the particle mass at low values of transverse 
momentum ($p_{\rm{T}} < 2$~GeV/$c$). This dependence is induced by the system's collective radial expansion, 
which being cumulative throughout the whole collision, has a significant contribution from both the partonic 
and the hadronic phases. Figure~\ref{fig3} presents the $p_{\rm{T}}$ differential $v_2$ for central Pb--Pb collisions 
(i.e. 10--20$\%$ centrality) for charged $\pi$, K, and $\bar{\rm{p}}$, $\rm{K}^0_S$ and $\Lambda$. The measurement was performed with the 
scalar product method applying a pseudo--rapidity gap of $\Delta\eta > 1$ to suppress correlations not related 
with the symmetry plane (i.e. non--flow effects). It is seen that the data points exhibit the characteristic mass ordering 
in the low $p_{\rm{T}}$ region (i.e. $p_{\rm{T}} \le 2$~GeV/$c$), attributed to collective radial flow that shifts heavy 
particles to higher values of $p_{\rm{T}}$. The experimental points are compared to hydrodynamical calculations 
\cite{Ref:Hydro} using a value of $\eta/s = 0.2$ and the Color Glass Condensate (CGC) initial conditions, represented 
by the solid curves. The model reproduces the $p_{\rm{T}}$ dependence of $v_2$ for pions and kaons up to 
$p_{\rm{T}} \approx 2$~GeV/$c$ but overestimates $v_2$ for heavier particles. 

A similar analysis has recently been performed in p--Pb collisions at $\sqrt{s_{NN}} = 5.02$~TeV 
and reported by ALICE  \cite{Ref:AlicepPb}. It was based on studying two--particle correlations measuring particle pairs from same events 
separated in the pair angles $\Delta\eta$ and $\Delta\varphi$ corrected for acceptance effects and efficiency 
losses using mixed events. The results revealed a double--ridge structure for $\pi$--hadron, 
K--hadron and p--hadron correlations after subtracting the associated yield per trigger particle in $\Delta\eta$ 
and $\Delta\varphi$ for the 60--100$\%$ multiplicity class from the corresponding yield from events with 20$\%$ 
highest multiplicity p--Pb collisions. This subtraction procedure was performed to reduce the contribution from 
the dominant near side jet structure. A selection of $\Delta\eta > 0.8$  was applied to reduce any residual contribution from the near side jet peak. The resulting distribution was projected on the $\Delta\varphi$ axis and was fitted with a Fourier series.
Figure \ref{fig4} presents the $p_{\rm{T}}$ differential $v_2$, 
which is the dominant harmonic, for $\pi$, K, p and charged particles. A mass ordering between $\pi$ and p is observed with a hint that this ordering is also followed by 
K at low $p_{\rm{T}}$ values. The resemblance with the corresponding Pb--Pb picture (see fig.~\ref{fig3}), where this ordering is attributed to collective effects, 
is striking.

\vspace{-0.4 cm}
\section{Two-particle correlations and their charge dependence in Pb--Pb}

\begin{figure}[h]
\begin{minipage}{17pc}
\includegraphics[width=17pc]{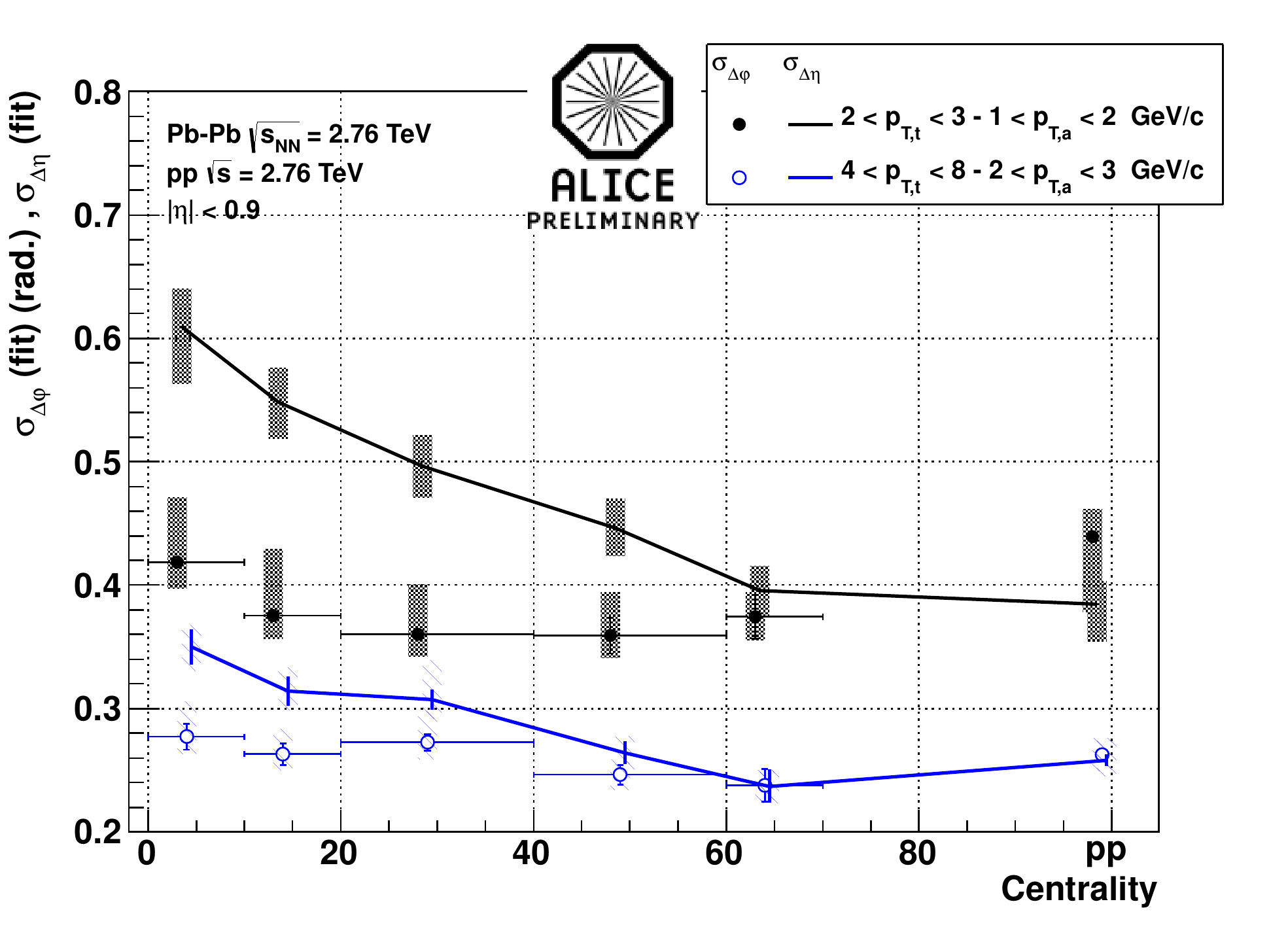}
\caption{\label{fig5}The centrality dependence of the width of the near--side jet--peak in $\Delta\eta$ and $\Delta\varphi$ for different ranges of $p_{\rm{T}}$ for the trigger and the associated particles, in Pb--Pb collisions.}
\end{minipage} 
\begin{minipage}{17pc}
\includegraphics[width=17pc]{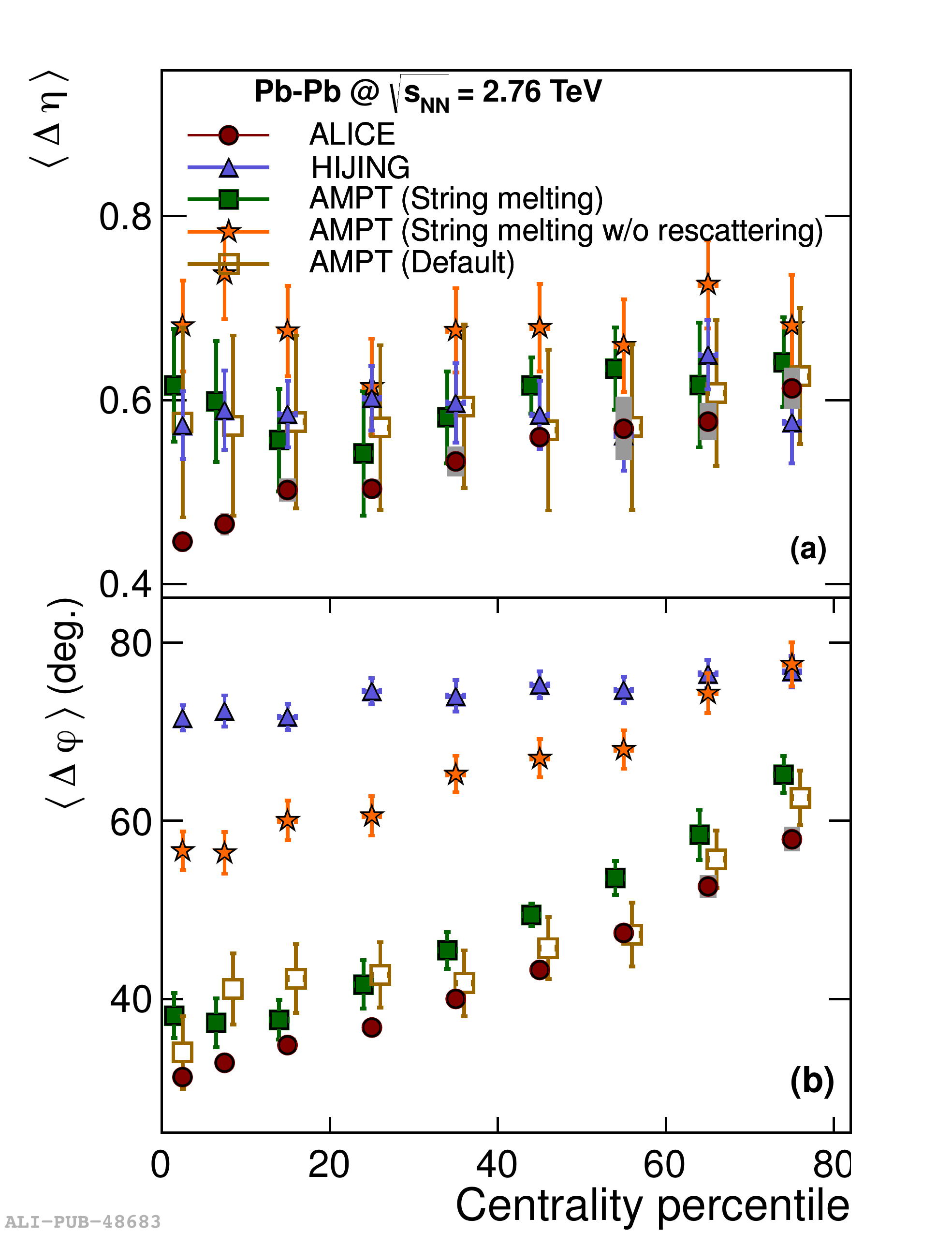}
\caption{\label{fig6}The centrality dependence of the width of the balance functions, $\langle\Delta\eta\rangle$ and $\langle\Delta\varphi\rangle$, in Pb--Pb collisions~\cite{Ref:AliceBF}.}
\end{minipage}\hspace{2pc}%
\end{figure}

The study of two--particle correlations described in the previous paragraph has been established as an 
important tool for the study of different momentum ranges e.g. the bulk, the jet dominated region, with different 
properties of the system being imprinted in the final observed structures. It was proposed in \cite{Ref:JetPeak} 
that using this type of analysis one has the possibility to probe the modification of the jet structure and shape, 
originating from the interplay with the longitudinally flowing medium. This was studied in Pb--Pb collisions, 
estimating the $\Delta\eta$ independent effects by quantifying the long range correlations on the near side 
(at $\Delta\eta > 1$), and subtracting them from the short range region (at $\Delta\eta < 1$). The resulting 
near side jet peak is then fitted with a double Gauss function. The resulting values are plotted as a function of centrality in fig.~\ref{fig5} for 
different values of $p_{\rm{T}}$ of the trigger and the associated particles. They show a significant centrality dependence for $\sigma_{\Delta\eta}$ but a rather mild effect for $\sigma_{\Delta\varphi}$. The data are 
described by AMPT calculations (not shown in the plot) and can be accommodated by models that consider the interplay 
of low $p_{\rm{T}}$ jets with the flowing medium.

The charge dependent part of these correlations has recently emerged as a key feature in many studies since 
they can probe the effect of local charge conservation coupled to a system that exhibits collective behaviour. This 
is addressed by the so called balance functions \cite{Ref:bfPratt}, defined as the difference between the unlike-- and like--sign 
pair densities divided by the number of trigger particles. Each pair density term is corrected 
for detector acceptance and efficiency effects using mixed events. The balance function measures the correlation 
strength of balancing partners, imprinted in the width of the distribution, and it is sensitive to the time of hadronization. Figure \ref{fig6} presents the centrality dependence in Pb--Pb collisions of the width of the 
balance function in $\Delta\eta$ ($\langle\Delta\eta\rangle$) and $\Delta\varphi$ ($\langle\Delta\varphi\rangle$), 
calculated as the weighted average from each distribution. It is seen that the width for data (red full circles) 
decreases for more central collisions in both representations, consistent with the picture of late stage creation of 
balancing charges focused by the strong radial flow. It is also interesting to note that only models that incorporate 
collective effects (e.g. AMPT) describe the data points in $\Delta\varphi$ but seem to fail to do so in 
$\Delta\eta$~\cite{Ref:AliceBF}.

\begin{figure}[h]
\begin{minipage}{17pc}
\includegraphics[width=17pc]{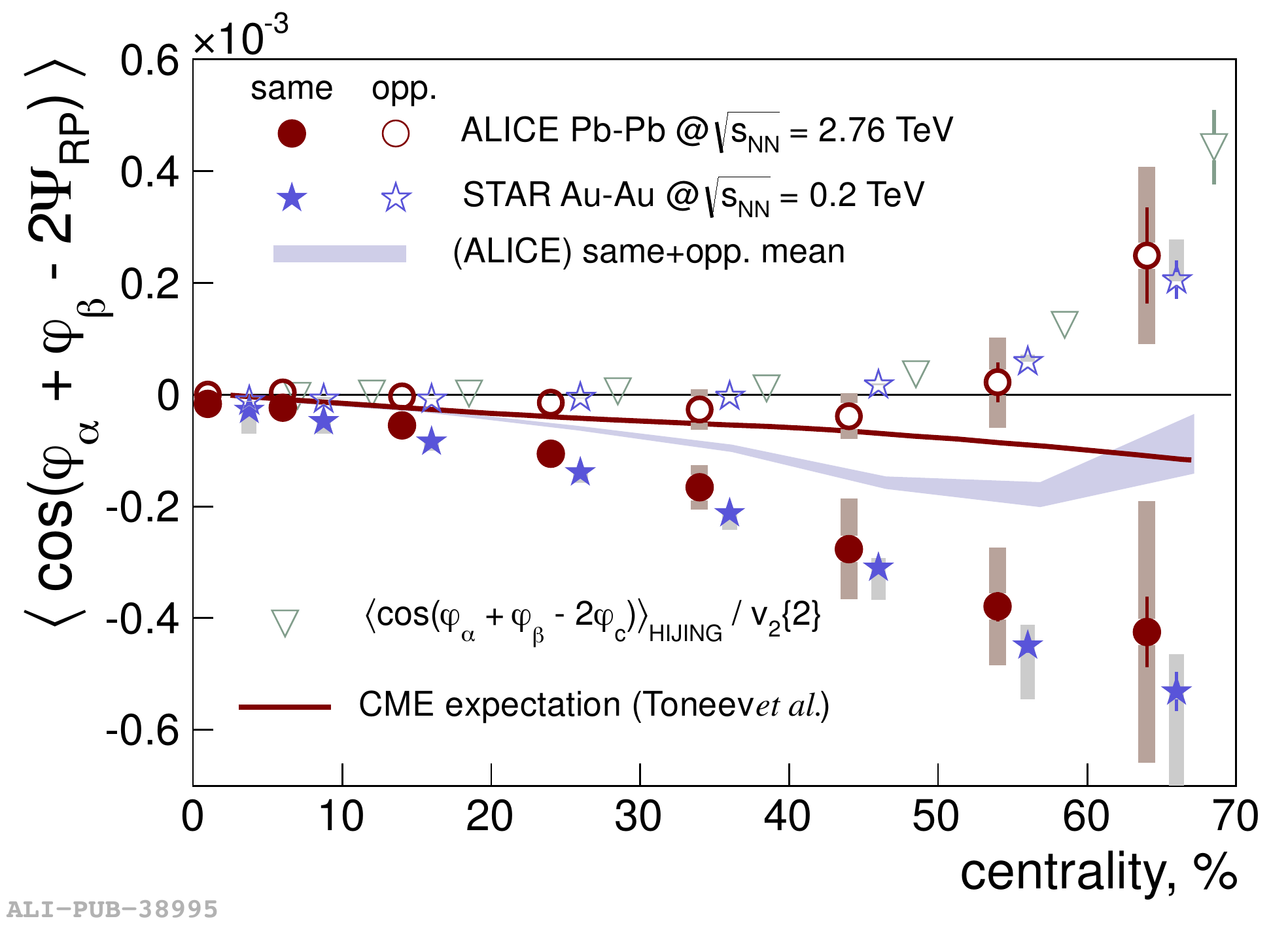}
\caption{\label{fig7}The centrality dependence of the charge dependent correlations of  $\langle\cos(\varphi_{\alpha} + \varphi_{\beta} - 2\Psi_{RP})\rangle$ in Pb--Pb collisions~\cite{Ref:AliceCME}.}
\end{minipage}\hspace{2pc}%
\begin{minipage}{17pc}
\includegraphics[width=17pc]{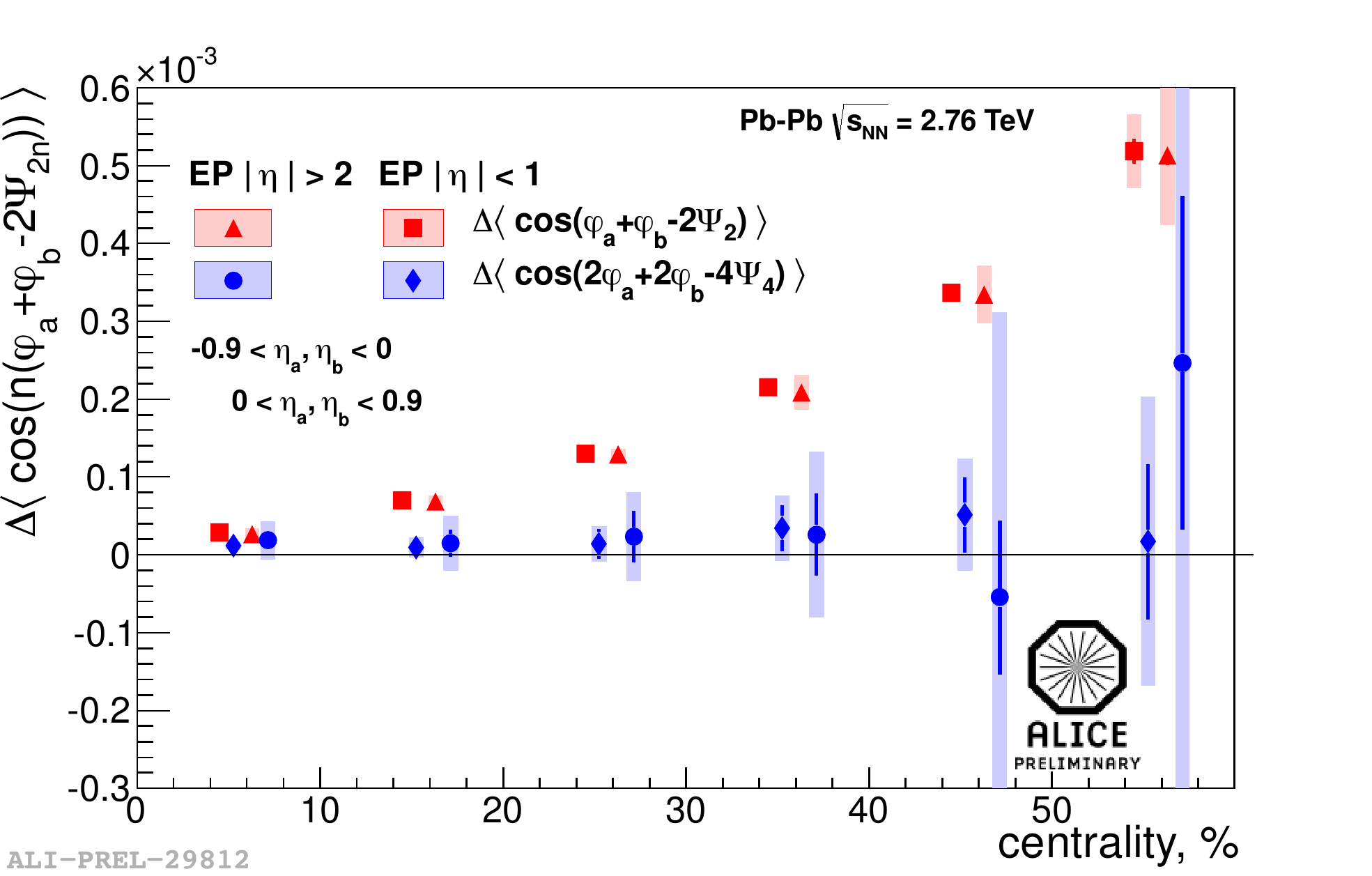}
\caption{\label{fig8}The centrality dependence of the differences of the charge dependent correlations of $\langle\cos(\varphi_{\alpha} + \varphi_{\beta} - 2\Psi_{2})\rangle$ and $\langle\cos(2\varphi_{\alpha} + 2\varphi_{\beta} - 4\Psi_{4})\rangle$ in Pb--Pb collisions.}
\end{minipage} 
\end{figure}

These charge--dependent correlations relative to the collision symmetry plane, were recently 
used to investigate the possibility of observing $P$-- and $CP$--odd effects in heavy--ion collisions. These 
effects originate from the interaction of local domains (i.e. QCD excited vacuum states) that can violate the $CP$ 
invariance, with the quarks of the deconfined phase and with the strong, of the order of $10^{19}$~G, magnetic 
field that develops in peripheral collisions at the LHC energies. This is better known as the Chiral Magnetic Effect 
(CME), the experimental consequence of which is a charge asymmetry relative to the reaction plane. The first 
results from both STAR~\cite{Ref:StarCME} and ALICE~\cite{Ref:AliceCME} are based on the study of correlations 
of two particle relative to the reaction plane (i.e. $\langle\cos(\varphi_{\alpha} + \varphi_{\beta} - 2\Psi_{RP})\rangle$) 
and show a strong charge dependent signal that develops for more peripheral events, qualitatively consistent with 
the CME expectations. This is illustrated in fig.~\ref{fig7} where the ALICE results for same and opposite charge 
combinations are represented by the red filled and open circles, respectively. It is seen that while the unlike--sign 
correlations are compatible with zero, with the exception of the most peripheral points, the like--sign ones exhibit 
an increasing magnitude for more peripheral events, coupled to a negative sign. It is also interesting to note that 
the LHC results are in quantitative agreement with the RHIC values (i.e. blue filled 
and open stars for the like and unlike--sign pairs, respectively) inpite the almost 14 times higher centre--of--mass 
energy at the LHC. 
The focus now turns to analyses whose aim is to disentangle the potential signal originating from the CME, 
from the background. For the latter, it was recently realised that the biggest contributor is the effect of local 
charge conservation coupled to elliptic flow modulations \cite{Ref:PrattCME}. One way to probe the background 
contribution is to study the charge dependent correlations relative to the fourth harmonic symmetry plane i.e. the 
so--called double harmonic correlator $\langle\cos(2\varphi_{\alpha} + 2\varphi_{\beta} - 4\Psi_{4})\rangle$. 
The contribution from the CME is argued to be little, if any at all (i.e. correlations relative to $\Psi_{4}$), while 
the background should create a charge dependent signal albeit scaled compared to the previous case (i.e. 
$\langle\cos(\varphi_{\alpha} + \varphi_{\beta} - 2\Psi_{2})\rangle$, the so--called single harmonic correlator) 
by the ratio of $v_4/v_2$. Figure~\ref{fig8} presents the centrality dependence of the differences between the 
correlations of unlike and like--sign pairs for both the single (i.e. red points) and the double harmonic (i.e. blue 
points) correlators. While the single harmonic correlator exhibits a rather strong charge dependent signal (see also fig.~\ref{fig7}), the double harmonic one is consistent with zero within the current large uncertainties. This could be interpreted as an indication that the effect of local charge 
conservation and its interplay with azimuthal anisotropy might not contribute  significantly in these measurements. However, before reaching a final conclusion, a measurement of better 
precision needs to be performed from the experimental side, while there is an increasing need for these effects 
to be studied thoroughly from the theory side e.g. using models that can describe both the hydrodynamical 
evolution of the system but also its final state, hadronic interactions. 

\vspace{-0.4 cm}
\section{Summary}
The study of angular correlations in Pb--Pb provided valuable information on the expansion of the system, driven by initial pressure gradients. During this phase it behaves as a liquid 
with low dissipation values. This opens up the possibility for the initial energy density fluctuations to be imprinted in 
the final state observables. These studies in p--Pb collisions have recently yielded surprisingly similar observations 
compared to the Pb--Pb results, the interpretation of which leads to rather intriguing paths.

\vspace{-0.5 cm}



\section*{References}
\begin{thebibliography}{9}
\bibitem{Ref:Heinz}  {U.~Heinz and R.~Snellings, Annu. Rev. Nucl. Part. Sci. \textbf{63}, 123 (2013).}
\bibitem{Ref:Alicev2}  {K.~Aamodt \textit{et al.} [ALICE Collaboration], Phys. Rev. Lett. \textbf{105}, 252302 (2010).}
\bibitem{Ref:LhcHigherHarmonics}  {K.~Aamodt \textit{et al.} [ALICE Collaboration], Phys. Rev. Lett. \textbf{107}, 032301 (2011).}
\bibitem{Ref:LhcHighPt}  {G.~Aad \textit{et al.} [ATLAS Collaboration], Phys. Rev. Lett. \textbf{105}, 252303 (2010); K.~Aamodt \textit{et al.} [ALICE Collaboration], Phys. Lett. \textbf{B696}, 30 (2011).}
\bibitem{Ref:AliceJinst}  {K. Aamodt \textit{et al.} [ALICE Collaboration], JINST \textbf{3}, S08002 (2008).}
\bibitem{Ref:Voloshin}  {S.~Voloshin and Y.~Zhang, Z. Phys. \textbf{C70}, 665 (1996).}
\bibitem{Ref:Alicev1}  {B.~I.~Abelev \textit{et al.} [ALICE Collaboration], arXiv:1306.4145 [nucl-ex].}
\bibitem{Ref:Rhicv1}  {B.~I.~Abelev \textit{et al.} [STAR Collaboration], Phys. Rev. Lett. \textbf{101}, 252301 (2008).}
\bibitem{Ref:TeanyYan}  {D.~Teaney and L.~Yan, Phys. Rev. \textbf{C83}, 064904 (2011).}
\bibitem{Ref:Hydro}  {C.~Shen \textit{et al.}, Phys.Rev. \textbf{C84}, (2011) 044903.}
\bibitem{Ref:AlicepPb}  {B.~I.~Abelev \textit{et al.} [ALICE Collaboration], arXiv:1307.3237 [nucl-ex].}
\bibitem{Ref:JetPeak}  {N.~Armesto \textit{et al.}, Phys. Rev. \textbf{C72}, 064910 (2005).}
\bibitem{Ref:bfPratt}  {S.~A.~Bass, P.~Danielewicz and S.~Pratt, Phys. Rev. Lett. \textbf{85}, 2689 (2000).}
\bibitem{Ref:AliceBF}  {B.~I.~Abelev \textit{et al.} [ALICE Collaboration], Phys. Lett. \textbf{B723}, 267 (2013).}
\bibitem{Ref:StarCME}  {B.~I.~Abelev \textit{et al.} [STAR Collaboration], Phys. Rev. Lett. \textbf{103}, 251601 (2009).}
\bibitem{Ref:AliceCME}  {B.~I.~Abelev \textit{et al.} [ALICE Collaboration], Phys. Rev. Lett. \textbf{110}, 012301 (2013).}
\bibitem{Ref:PrattCME}  {S.~Schlichting and S.~Pratt, Phys. Rev. \textbf{C83}, 014913 (2011).}
\end{thebibliography}
\end{document}